\documentclass[11pt]{article}
\begin{document}
\begin{center}
{\large\scshape
%%%%%%%%%%%%%%%%%%%%%
%% Title

Accretion Discs with an Inner Spiral Density Wave
}
\vspace{3mm}
\end{center}

\begin{center}
{  
%%%%%%%%%%%%%%%%%%%%%
%% Authors
M.M. Montgomery$^{1}$ \& D.V. Bisikalo$^{2}$

}
\end{center}

\begin{center}
\noindent
{\small \em
%%%%%%%%%%%%%%%%%%%%%
%% Affiliations
$^{1}$ Department of Physics, University of Central Florida, Orlando, FL  32816, USA\\
        $^{2}$ Institute of Astronomy, Moscow
}
\end{center}
{
%%%%%%%%%%%%%%%%%%%%%
%% Abstract
\begin{abstract}
In Montgomery (2009a), we show that accretion discs in binary systems could retrogradely precess by the same physics that causes the Earth to retrogradely precess [i.e., tidal torques by the Moon and the Sun (or secondary star) on a tilted, spinning, non-spherical Earth (or e.g., primary star surrounded by an accretion disc)].  In addition, we show that the state of matter and the geometrical shape of the celestial object could significantly affect the precessional value.  For example, a Cataclysmic Variable (CV) Dwarf Novae (DN) non-magnetic system that shows negative superhumps in its light curve can be described by a retrogradely precessing, differentially rotating, tilted disc.  Because the disc is a fluid and because the gas stream overflows the tilted disc and particles can migrate into inner disc annuli, coupled to the disc could be a retrogradely precessing inner ring that is located near the innermost annuli of the disc.  However, numerical simulations by Bisikalo et al. (2003, 2004) and this work show that an inner spiral density wave can be generated instead of an inner ring.   Therefore, we show that retrograde precession in non-magnetic, spinning, tilted CV DN systems can equally be described by a retrogradely precessing and differentially rotating disc with an attached retrogradely precessing inner spiral density wave so long as the wave appears at the same radius as the ring and within the plane of the tilted disc. We find that the theoretical results generated in this work agree well with the theoretical results presented in Montgomery (2009a) and thus with the numerical simulations and select CV DN systems in Montgomery (2009b) that may have a main sequence secondary.  Therefore, pressure effects do need to be considered in CV DN systems that exhibit negative superhumps if the accretion discs are tilted and have an inner spiral density wave that is in the plane of the disc.
\end{abstract}


\section{Introduction}
Variability in non-magnetic, Cataclysmic Variable (CV) close binaries has been largely attributed to instabilities.  The outbursts in CV recurrent novae (RN) and CV classical novae (CN) can be caused by thermonuclear runaway on the surface of a white dwarf or by a thermal and viscous instability in the disc.  The latter type of outburst involves cycling from low-to-high thermal and viscous states (see e.g., Meyer \& Meyer-Hofmeister 1981 and references within), and thus cycling from low-to-high mass transfer rates (see e.g., Lasota 2001), that results in a release of accretion energy as mass is drained onto the white dwarf.  Transient X-ray binaries with neutron stars or black hole primaries and CV Dwarf Novae (DN) are also subject to thermal and viscous instabilities.  However, the thermal and viscous instabilities in CV RN are larger in scale than those in CV DN.  In addition to thermal and viscous instabilities, CV DN can experience a tidal instability.  Results are CV DN SU UMa and Nova-Like (NL) outbursts.  The larger amplitude and longer duration superoutbursts involve both a thermal-viscous instability and a tidal instability.  

Both CV DN SU UMa and NL outbursts and superoutbursts show characteristic hump-shaped modulations, known as positive superhumps, in their light curves.  These positive superhump modulations have a period that is a few percent longer than the orbital period.  They could be caused by dissipation from spiral shocks that formed from converging gas flows in the middle and outer annuli of the disc (see e.g., Smith et al. 2007).  The source to this spiral shock generation is the \( j : j - 1 = 3 : 2 \) resonance (Lubow 1991a,b), using the terminology established by Frank, King, \& Raine (1992).  A competing theory has also been suggested by Bisikalo et al. (2004) - the superoutburst is associated with the generation of an spiral density wave in inner annuli of the disc, a zone that is denser and gas-dynamically un-perturbed.  This inner spiral density wave contributes to a  substantial increase in mass transfer rate in that region.

In addition to positive superhumps, some non-magnetic CV DN systems show negative superhumps in their light curves.  These negative superhump modulations have a period that is a few percent shorter than the orbital period.  Some systems' light curves only show negative superhumps (e.g., CN Ori), or only show positive superhumps (e.g., SU UMa), or show both modulations simultaneously [e.g., SDSS J210014.12+00446.0 (Olech et al. 2009)], or show one type of modulation only later to be replaced by the other [e.g., TT Ari (Kim et al. 2009)].  Negative superhumps are suggested to be a consequence of a partially tilted disc (Patterson et al. 1993) or are related to a warped accretion disc (Petterson 1977; Murray \& Armitage 1998; Terquem \& Papaloizou 2000; Murray et al. 2002; Foulkes, Haswell, \& Murray 2006).  Our previous numerical simulation work (Wood, Montgomery, \& Simpson 2000; Montgomery 2004, Montgomery 2009b) shows that negative superhumps can be produced by a fully tilted disc, and these modulations have a shape and a period that are similar to those in observational light curves.

Accretion discs can warp or tilt via a potpourri of suggested sources.  For example, disc tilt in X-ray binaries can be from gas streaming at an upward angle from the inner Lagrange point for one half of the orbit and at a downward angle for the second half of the orbit (Boynton et al. 1980).  In some CVs, a disc tilt can be held constant by gas stream being fed by the magnetic field of the secondary (Barrett, O'Donoghue, \& Warner, 1988).  Also for CV systems, a disc tilt instability can result from a coupling of an eccentric instability to Lindblad resonances (Lubow 1992).  A vertical resonant oscillation of the disc mid-plane can be caused by tidal interactions between a massive secondary and a coplanar primary (Lubow \& Pringle 1993).  A warping instability can be caused by irradiation from the primary (Pringle 1996, 1997).  A warping can be caused by direct tidal forces from a secondary that is on an inclined orbit.  A disc warp can be caused by misalignments of the spin axis of a compact and/or magnetized primary and the disc axis.  Of these possibilities, Murray \& Armitage (1998) find that CV DN accretion discs do not seem to tilt out of the orbital plane by instabilities.  

If a warp is present, the warp may or may not propagate. As reviewed in Pringle (1999), an inviscid, locally isothermal or locally polytropic fluid disc that has Keplerian rotation, that is within a spherically symmetric gravitational potential, and that is wholly tilted by the same finite angle has azimuthal wavenumber $m$=1[and thus the fluid disc has ($m$=1)-fold symmetry in the azimuthal direction] and radial mode number $n$=0 (and thus the fluid disc has $n$=0 local minima).  For this disc, the wave mode frequency is zero.  If, however, not all of the inviscid disc is tilted by the same amount, then a wave can radially propagate (see Papaloizou \& Lin, 1995 and e.g., Lubow, Ogilvie, \& Pringle, 2002).  If the disc has some viscosity, then a disc wave can only propagate if the viscosity is low and the tilt angle is approximately less than $H/r$ where $H$ is vertical disc thickness and $r$ is disc radius.  If the disc has low viscosity, then the wave is transferred through the disc by wave action.  Likewise, if the disc has high viscosity, then the wave is transferred through the disc by dispersion (Papaloizou \& Pringle, 1983).  If a quiescent, viscous disc is wholly tilted by the same amount and the tilt angle is greater than $H/r$, a wave should not propagate. 

Regardless of how an accretion disc tilts or warps and whether the wave propagates, non-coplanar accretion discs in binary systems precess in the retrograde direction (i.e., in the direction opposite to spin and orbital motion).  Retrograde precessional activity has been suggested for X-ray binaries (e.g., Katz 1973, Larwood 1998, Wijers \& Pringle 1999), for common systems that produce jets (e.g., Livio 1999), for protostars (e.g., Papaloizou \& Terquem 1995, Larwood et. al. 1996, Larwood 1997), for CV systems (Bisikalo et al. 2003, 2004) and for quasars and supermassive black holes (e.g., Romero et al. 2000, Caproni \& Abraham 2002), to name a few.  In many of these binary systems, the source to the retrograde precession is often suggested to be due to direct tidal forcing.  In Montgomery (2009a), we show that tidal torques by the secondary on a tilted, spinning accretion disc is a source to retrograde precession.  A more familiar example is the tidal torque by the Sun and Moon on the oblate, spinning, tilted Earth that causes the Earth's First Point of Aries to precess retrogradely.  

Of note is that some find retrograde precession occurring within a disc even though the entire accretion disc remains coplanar with the orbit:  In numerical simulations by Bisikalo et al. (2003, 2004), a coplanar inner spiral density wave is generated within the inner annuli of a coplanar accretion disc and only the coplanar spiral density wave retrogradely precesses.  Based on their numerical simulations, the wave is generated by sheared elliptical orbits and these sheared elliptical orbits (and thus the wave) retrogradely precess.  Bisikalo et al. (2004) suggest that the retrograde precession is due to an influence by the mass-losing star by the processes discussed in Kumar (1986) and Warner (1995, 2003).  However, Kumar (1986) finds the retrograde precession on a tilted disc's $\emph{outer}$ annulus due to the tidal influence of the secondary.  Since the spiral density wave is found in inner annuli of the accretion disc, theory is needed on retrograde precession for a wave in this location.   

In Montgomery (2009a), we suggest that care must be taken in choosing the correct state of matter and the correct geometrical shape of the accretion disc as different shapes can produce different retrograde precessional values.   For example, Montgomery (2009a) shows that retrograde precession of a non-magnetic, tilted CV DN gaseous accretion disc can best be described by a retrogradely precessing and differentially rotating tilted accretion disc, and superimposed on the disc is a rotating and precessing ring that is located near the innermost annuli of the disc.  The superimposed ring is likely caused by gas stream overflow and additional particle migration into the innermost annuli of the tilted disc (Montgomery 2009a,b). Instead of a precessing inner ring in accretion discs, some accretion disc simulations show an inner spiral density wave in the same location:  For semidetached binaries, Bisikalo et al. (2003, 2004) show a one-armed spiral shock wave is generated in inner annuli of cooler discs (i.e., $T \sim 10^{4} K$) of binaries with high secondary-to-primary mass ratios.  Therefore, a study is needed on whether such a wave can be generated in simulations of lower mass ratio, and can such a wave explain retrograde precession in CV DN systems.  

In this work, we generate theory on retrograde precession of an accretion disc that contains an inner spiral density wave.  We assume the wave is within the plane of a quiescent disc, yet the disc is tilted five degrees relative to the orbital plane.  We wish to establish whether replacing the precessing inner ring found in Montgomery (2009a) with an inner spiral density wave has similar effects on disc retrograde precession.  To justify the disc geometry (i.e.. disc with a coplanar inner spiral density wave), we generate a low mass ratio numerical simulation using the grid code of Bisikalo et al. (2003).  In \S 2 of this paper, we develop the theoretical expressions to describe a retrogradely precessing, tilted CV DN accretion disc that has an inner spiral density wave in the plane of the disc.  In \S 3, we describe the two numerical codes that generate retrograde precession in CV DN accretion discs and we present numerical simulation results.  In \S 4, we compare our theoretical and numerical results with results obtained from observations and with previously obtained results.  In \S 5, we discuss these comparisons, and in \S 6, we provide a summary, conclusions, and future work.

\section{Theoretical Expressions for Retrograde Precessing of Discs with an Inner Spiral Density Wave}
In the Montgomery (2009b) Smoothed Particle Hydrodynamic (SPH) numerical simulations of a differentially rotating, retrogradely precessing, tilted accretion disc, a ring forms near the innermost annuli of the disc due to particle migration into this inner annuli and due to gas stream overflow of the tilted disc edge.   Instead of a ring, numerical simulations by Bisikalo et al. (2003. 2004) show gas stream build-up in the form of a one-armed spiral density wave.   Therefore, we consider a hypothetical disc geometry for non-magnetic CV DN system that exhibits negative superhumps in their light curves.  The hypothetical disc geometry is a combination of the disc geometries in both numerical simulations:  We assume the disc is wholly tilted five or more degrees degrees relative to the orbital plane.  Instead of a denser ring of material in the innermost annuli of the disc, we assume an inner spiral density wave that remains in the plane of the tilted disc.  In the following subsections, we find the disc's net retrograde precessional value. 

\subsection{Retrograde Precession, Dynamical Term}
The retrograde precessional angular frequency of a differentially rotating, circular, tilted disc is found to be (see e.g., Montgomery, 2009a) 
%%%%%%%%%%%%%%%EQUATION 1%%%%%%%%%%%%%%%%%%%%%%%%
\begin{eqnarray}
\omega_{r,dyn} & = & - \frac{15}{32} \frac{q}{\sqrt{(1+q)}} \cos\theta \left( \frac{r_{d}}{d} \right)^{3/2} \omega_{orb}.
\end{eqnarray}
%%%%%%%%%%%%%%%%%%%%%%%%%%%%%%%%%%%%%%%%%%%%%%

\noindent
In this equation, $q=M_{2}/M_{1}$ is the mass ratio of the secondary-to-primary masses, $\theta$ is the obliquity angle, $r_{d}$ is the radius of the disc, $d$ is the orbital separation, and $\omega_{orb}$ is the orbital frequency.  This equation applies to a thin disc that lacks an inner hole or has a small inner hole (e.g., non-magnetic, CV DN systems).  This disc also does not have any surface features such as spiral density waves or dense rings (or winds, jets, planets, holes, gaps, or any other feature that may change the angular momentum, and thus moment of inertia, of the system).  The shape of the disc is ideally circular although the equation is assumed applicable to non-circular discs as well (see Montgomery 2009a).  Equation (1) is valid so long as the disc precesses as a unit and the distance between the stars $d$ is much greater than the radius of the primary (see Montgomery 2009a).  

The outer radius of the approximately circular disc is assumed to be

%%%%%%%%%%%%%%%%%EQUATION 2%%%%%%%%%%%%%%%%%%%%%%%
\begin{equation}
r_{d}=0.6d/(1+q)
\end{equation}
\noindent
%%%%%%%%%%%%%%%%%%%%%%%%%%%%%%%%%%%%%%%%%%%%%%%
\noindent
as given by Paczynski (1977).  This equation is valid for mass ratios in the range \( 0.03<q<1 \).  

As retrograde precession is associated with tilted discs and as negative superhumps are generated when disc tilt is high enough (see Montgomery 2009b), then negative superhumps could be an observational beacon that indicates that the disc may be tilted.  As the negative superhump period $P_{-}$ and the orbital period $P_{orb}$ are fairly precise clocks, we would like to relate the retrograde precessional period to these precise clocks.  From observations, we know that the orbital period is slightly longer than the negative superhump period.  Both periods are related to the net retrograde precession period by 

%%%%%%%%%%%%%%%%%%EQUATION 3%%%%%%%%%%%%%%%%%%%%%%
\begin{equation}
P_{r,net}^{-1} = P_{-}^{-1} - P_{orb}^{-1}.  
\end{equation}
%%%%%%%%%%%%%%%%%%%%%%%%%%%%%%%%%%%%%%%%%%%%%%%

\noindent
If we assume a primary mass $M_{1}$=0.8$M_{\odot}$ and mass ratio $q=0.4$, then the secondary mass $M_{2}$=0.32$M_{\odot}$, $r_{d}\sim0.43d$ where $d\sim1.23R_{\odot}$ (Montgomery 2009b).  From Newton's version of Kepler's third law, the orbital period and the rotation period at the disc edge is $P_{orb}=3.852$ hours and one hour, respectively.  From Montgomery (2009b), the negative superhump period for this mass ratio is $P_{-} \sim$ 3.662 hours and therefore, from Equation (3), $P_{r,net}\sim$19.5$P_{orb} \sim $75 hours, assuming small angle approximation for disc tilt.  Notice that the spin rate at the edge of the disc is fast compared to the retrograde precessional rate.  However, this type of precession is not fast precession because the wobble rate is not comparable to the spin rate.  Instead, this kind of precession is driven by torques (see e.g., Montgomery 2009a).  

\subsection{Retrograde Precession, Pressure Term}
To find the effects an inner spiral density wave has on retrograde precession, we first review the  theory generated for CV DN accretion discs that progradely precess, that have spiral density waves, and that generate positive superhumps.  From the location of the spiral density waves in a progradely precessing disc, we can find the location of the inner spiral density wave.

\subsubsection{Review:  Net Prograde Precession in CV DN}
Spiral density waves contribute to retrograde precession through pressure as Lubow (1992) finds  for progradely precessing CV DN accretion discs that show positive superhumps in their light curves.  Lubow (1992) finds that the net prograde precession $\omega_{p,net}$ is the net of a dynamical component, a pressure component, and a transient component.  Of the three, the dynamical component contributes most to prograde precession.  The second of the three is a retrograde component and is a correction term to the dynamical component.  This pressure component is necessary to more correctly explain the net prograde precession in steady state discs.  The third component is also a correction term and can contribute to the prograde dynamical component or to the retrograde pressure component.  However, this transient component is most effective during the development stage of positive superhumps.  If only quasi-steady state discs are considered, then this third component can be ignored.  

Pressure effects within progradely precessing discs could be due to a two-armed spiral mode.  The source to this spiral shock generation is the \( j : j - 1 = 3 : 2 \) resonance (Lubow 1991a,b), using the terminology established by Frank, King, \& Raine (1992).   Dissipation from these spiral shocks are created from converging gas flows, and this dissipation is the source to the observed positive superhump as suggested by Smith et al. (2007).  The location of these spiral density waves is in the middle and outer annuli of the quasi-steady state disc (see e.g., Smith et al., 2007).  

Assuming this disc geometry for prograde precession, an analytical expression for the pressure component is developed in Lubow (1992).  The standard Wentzel-Kramers-Brillouin (WKB) relation for spiral density waves in the tight winding limit is given by 
%%%%%%%%%%%%%%%%EQUATION 4%%%%%%%%%%%%%%%%%%%%%%%%%%
\begin{equation}
(\Omega' - \omega_{press,outer})^{2} = \kappa'^{2} + k'^{2}c^{2}
\end{equation}
%%%%%%%%%%%%%%%%%%%%%%%%%%%%%%%%%%%%%%%%%%%%%%%%

\noindent
where $\Omega'$ and $\kappa'$ are the particle angular and radial frequencies, respectively, $\omega_{press,outer}$ is the oscillation frequency of the spiral density waves, $c$ is the gas sound speed within the disc, and $k'$ is the radial wavenumber of the spiral arms (see e.g., Goldreich \& Tremaine 1979).  In a non-precessing disc, $\Omega' = \kappa'$, and in a slowly precessing disc, $\Omega' \approx \kappa'$.  In the limit where $\omega_{press,outer} \ll \Omega'$ or $\kappa'$ and applying the previous approximation, Lubow (1992) approximates $\omega_{press,outer}$ in the previous equation.  Then Lubow (1992) couples $\omega_{press,outer}$ with the dynamical motion within the disc to find the net prograde precession:
%%%%%%%%%%%%%%%%EQUATION 5 and 6 %%%%%%%%%%%%%%%%%%%%%%
\begin{eqnarray}
\omega_{p,net} 	& = & \omega_{p,dyn} - \frac{k'^{2}c^{2}}{2\Omega'}  \\
		                            & = & \omega_{p,dyn} - \omega_{press,outer}.
\end{eqnarray}
%%%%%%%%%%%%%%%%%%%%%%%%%%%%%%%%%%%%%%%%%%%%%%%%

\noindent
We identify the last term in the equation as the pressure term associated with these spiral density waves located in the middle-to-outer portion of the disc.  Note that we assume quasi-static discs; the net prograde precession has only a dynamical prograde component and a retrograde pressure component.

In e.g., Montgomery (2001), we show that the positive superhump period excess is 
%%%%%%%%%%%%%%%%EQUATIONS 7, 8, 9%%%%%%%%%%%%%%%%%%%
\begin{eqnarray}
\epsilon_{+} & = & \frac{P_{+}}{P_{orb}} - 1 \\
             & = & \left[ 1 - \frac{P_{orb}}{P_{p,net}} \right]^{-1} - 1 \\
             & = & \frac{P_{orb}}{P_{p,net} - P_{orb}} 
\end{eqnarray}

\noindent
where we have substituted  

%%%%%%%%%%%%%%%EQUATION 10%%%%%%%%%%%%%%%%%%%%%%%%
\begin{equation}
P_{p,net}^{-1} = P_{orb}^{-1} - P_{+}^{-1}
\end{equation}
%%%%%%%%%%%%%%%%%%%%%%%%%%%%%%%%%%%%%%%%%%%%%%

\noindent
into Equation (7) to obtain Equation (8).  In these equations, $P_{+}$ is the positive 
superhump period and $P_{p,net}$ is the net prograde precessional period.  If we assume \( \omega_{p,net} \ll \omega_{orb} \), then we obtain the positive superhump period excess
%%%%%%%%%%%%%%%%EQUATIONS 11 & 12%%%%%%%%%%%%%%%%%%%
\begin{eqnarray}
\epsilon_{+} & \sim & \frac{\omega_{p,net}}{\omega_{orb}} \\
	     & \sim & \frac{\omega_{p,dyn}}{\omega_{orb}} - \frac{\omega_{press,outer}}{\omega_{orb}}.
\end{eqnarray}
%%%%%%%%%%%%%%%%%%%%%%%%%%%%%%%%%%%%%%%%%%%%%

\noindent
The pitch angle $i$ of the spiral wave is related to the wavenumber by

%%%%%%%%%%%%%%%%%EQUATION13%%%%%%%%%%%%%%%%%%%%%%%%
\begin{equation}
\cot i' = k'r'
\end{equation}
%%%%%%%%%%%%%%%%%%%%%%%%%%%%%%%%%%%%%%%%%%%%%%%%
\noindent
where $r'$ is the radial distance to that fluid element within a spiral arm.  

In Montgomery (2001), we assume average values for $c$ and $i'$ based on our numerical simulations.  We find the average distance to the two numerically generated spiral density arms and the average pitch angle to be $r' \sim 0.477d$ and $i' \sim 17^{\circ}$, respectively, where $d$ is the binary separation.  After substitution of $r'$, $i'$, $k'$, $\Omega' = 3\omega_{orb}$, and the Smith \& Dhillon (1998) secondary mass-period relation into Equations (5) and (6), we find
%%%%%%%%%%%%%%%%%EQUATION 14%%%%%%%%%%%%%%%%%%%%%%%
\begin{eqnarray}
\omega_{press,outer} & = & \omega_{orb}\left[\frac{q^{0.422}}{28.4(1+q)^{2/3}+q^{0.422}}+1\right]^{-1} \\ 
                                   & \sim & \omega_{orb}\left[\frac{28.4(1+q)^{2/3}}{q^{0.422}}\right]^{-1} \nonumber.
\end{eqnarray}
%%%%%%%%%%%%%%%%%%%%%%%%%%%%%%%%%%%%%%%%%%%%%%%

\subsubsection{Retrograde Precession Effects from Inner Spiral Density Wave}
As reviewed in Kato (2001), local analytical studies of hydrodynamic accretion discs have found three types of trapped global modes.  One of the three types is inertial acoustic $\emph{p}$-modes (the other two are gravity $\emph{g}$-modes and vertical$\emph{c}$-modes and are not discussed in this work).  For a two-dimensional disc with no vertical modes, the $\emph{p}$-mode is described by 
%%%%%%%%%%%%%%%%EQUATIONS 15 %%%%%%%%%%%%%%%%%%%
\begin{equation}
(m\Omega'' - \omega_{press,inner})^{2} = \kappa''^{2} + k''^{2}c^{2}
\end{equation}
%%%%%%%%%%%%%%%%%%%%%%%%%%%%%%%%%%%%%%%%%%%%%
\noindent
where $\Omega''$ and $\kappa''$ are the particle angular and radial frequencies, respectively, $\omega_{press,inner}$ is the oscillation frequency of an inner spiral density wave, $c$ is the gas sound speed within the disc, $k''$ is the radial wavenumber of the spiral arms, and $m$ is an integer that indicates the different azimuthal modes.  Notice that this equation is very similar to Equation (4), however this equation is evaluated at the radius $r''$ of inner spiral density wave and hence the wave number and particle angular and radial frequencies are different.  Also note that we assume the same sound speed throughout the disc.  

Global spirals in discs are the $m=1$ $\emph{p}$-modes.  For a one-armed spiral wave, we can reduce the previous equation to 
%%%%%%%%%%%%%%%%EQUATIONS 16 %%%%%%%%%%%%%%%%%%%
\begin{equation}
\omega_{press,inner} = \Omega'' \pm \sqrt{\kappa''^{2} + k''^{2}c^{2}}.
\end{equation}
%%%%%%%%%%%%%%%%%%%%%%%%%%%%%%%%%%%%%%%%%%%%%
\noindent
In a non-precessing disc, $\Omega'' = \kappa''$, and in a slowly precessing disc, $\Omega'' \approx \kappa''$.  If the disk is also geometrically thin and $ck''\ll\Omega''$ (i.e., we have a global oscillation), then the previous equation reduces to
%%%%%%%%%%%%%%%%EQUATIONS 17 %%%%%%%%%%%%%%%%%%%
\begin{equation}
\omega_{press,inner}  \approx - \frac{k''^{2}c^{2}}{2\Omega''}.
\end{equation}
%%%%%%%%%%%%%%%%%%%%%%%%%%%%%%%%%%%%%%%%%%%%%
\noindent
This result is similar to that found by Kato (1983).  As Kato (1983) noted, this oscillation at r'' moves much slower than the angular velocity of the disc rotation.  Therefore, the oscillation acts to slow the motion of the gas particles in the disc and hence the sign is negative in this equation.  

To find the net retrograde precession in the disc, we need to sum the effects due to the inner spiral density wave with the effects due to the dynamical motions in the smooth, featureless disc: 
%%%%%%%%%%%%%%%%EQUATIONS 18 and 19%%%%%%%%%%%%%%%%%%%
\begin{eqnarray}
\omega_{r, net}  & = & \omega_{r,dyn} - \frac{k''^{2}c^{2}}{2\Omega''} \\
                             &  = & \omega_{r,dyn} - \omega_{press,inner}
\end{eqnarray}
%%%%%%%%%%%%%%%%%%%%%%%%%%%%%%%%%%%%%%%%%%%%%%
\noindent
The last term in the equation is identified as the pressure term associated with the inner spiral density wave number $k''$ and particle angular frequency $\Omega''$.  Notice that these equations are very similar to Equation (5) and (6), however these equations are evaluated at at the radius of the inner spiral density wave $r''$.  

In Montgomery (2009b), we show that the negative superhump period excess is 
%%%%%%%%%%%%%%%%EQUATIONS 20, 21, 22%%%%%%%%%%%%%%%%%%%
\begin{eqnarray}
\epsilon_{-} & = & 1 - \frac{P_{-}}{P_{orb}} \\
             & = & 1- \left[ 1 + \frac{P_{orb}}{P_{r,net}} \right]^{-1}  \\
             & = & \frac{P_{orb}}{P_{r,net} + P_{orb}}
\end{eqnarray}
%%%%%%%%%%%%%%%%%%%%%%%%%%%%%%%%%%%%%%%%%%%%%%
\noindent
where we have substituted Equation (3) into Equation (20) to obtain Equation (21).  In these equations, $P_{-}$ is the negative superhump period and $P_{r,net}$ is the net retrograde precessional period.  If we assume \( \omega_{p,net} \ll \omega_{orb} \), then we obtain the net negative superhump period excess 
%%%%%%%%%%%%%%%%EQUATIONS 23 & 24%%%%%%%%%%%%%%%%%%%
\begin{eqnarray}
\epsilon_{-,net} & \sim & \frac{\omega_{r,net}}{\omega_{orb}} \\
	     & \sim & \frac{\omega_{r,dyn}}{\omega_{orb}} - \frac{\omega_{press,inner}}{\omega_{orb}}.
\end{eqnarray}
%%%%%%%%%%%%%%%%%%%%%%%%%%%%%%%%%%%%%%%%%%%%%
\noindent
As $\omega_{r, dyn}$ is negative, then $\omega_{r, net}$ and $\epsilon_{-, net}$ are also negative.  If we assume a secondary mass-period relationship by Smith \& Dhillon (1998), then we can find the net negative superhump period excess in terms of orbital period (not shown).

The pitch angle of the inner spiral density wave $i''$ is related to wavenumber $k''$ by 
%%%%%%%%%%%%%%%%%EQUATION 25%%%%%%%%%%%%%%%%%%%%%%%%
\begin{equation}
\cot i'' = k''r''.
\end{equation}
%%%%%%%%%%%%%%%%%%%%%%%%%%%%%%%%%%%%%%%%%%%%%%%%
\noindent
Upon substitution into Equations (18) and (19), 
%%%%%%%%%%%%%%%%%EQUATIONS 26 & 27%%%%%%%%%%%%%%%%%%%%
\begin{eqnarray}
\omega_{press,inner}  & = & \frac{c^{2}}{2\tan^{2}i\Omega'' r''^{2}}  \\
                                         & \sim & \frac{19.53c^{2}}{\tan^{2}i\Omega'' d^{2}} 
\end{eqnarray}
\noindent
%%%%%%%%%%%%%%%%%%%%%%%%%%%%%%%%%%%%%%%%%%%%%%%%
where in the second version of $\omega_{press,inner}$ we have assumed r''$\sim$0.16d for the average outer radius of the inner spiral density wave.  This location is the average outer radius of the inner ring seen in the $q=0.4$ simulation of Montgomery (2009a) and is larger than $r_{circ}\sim$0.13d which is the minimum radius of the ring's outer radius.  

If $d$ is obtained from Kepler's Third Law, $c\sim$20 kms$^{-1}$ and $i'' \sim 17^{\circ}$ (Montgomery 2001), then only the particle orbital frequency $\Omega''$ is unknown.  However, we can easily find the particle orbital frequency through Kepler's Laws:  If we solve Newton's version of Kepler's third law for the orbital period and ratio it with Newton's version of Kepler's third law for the period of a particle within the disc $P$ at any radius $r$, ignoring perturbation effects by the secondary, then we find

%%%%%%%%%%%%%%%%%%EQUATION 28%%%%%%%%%%%%%%%%%%%%
\begin{equation}
\frac{P}{P_{orb}} = \left(\frac{r}{d}\right)^{3/2} (1+q)^{1/2}.
\end{equation}
%%%%%%%%%%%%%%%%%%%%%%%%%%%%%%%%%%%%%%%%%%%%%
\noindent 
After substitution for $r''$ and q, we find  \( P'' \sim 0.08 P_{orb} \) or \( \Omega''=12.5\omega_{orb} \).  We can use the Smith \& Dhillon (1998) secondary mass-period relation to find 
%%%%%%%%%%%%%%%%%%EQUATIONS 29%%%%%%%%%%%%%%%%%%%%
\begin{equation}
\omega_{orb}=2.54\times10^{-4} q^{-1/1.58}  \nonumber
\end{equation}
%%%%%%%%%%%%%%%%%%%%%%%%%%%%%%%%%%%%%%%%%%%%%
\noindent
where $\Omega_{orb}$ is in units of rad s$^{-1}$.  Upon substitution into Equations (26) or (27), we find
%%%%%%%%%%%%%%%%%EQUATIONS 30 & 31%%%%%%%%%%%%%%%%%%%
\begin{eqnarray}
\omega_{press,inner} & = & \omega_{orb}\left[145\frac{(1+q)^{2/3}}{q^{0.422}} \tan^2 i'' \right]^{-1} \\
                                        & = & \omega_{orb}\left[13.5\frac{(1+q)^{2/3}}{q^{0.422}}\right]^{-1},
\end{eqnarray}
%%%%%%%%%%%%%%%%%%%%%%%%%%%%%%%%%%%%%%%%%%%%%%%

\noindent
and \( \omega_{press,inner} \sim 2.1 \left( \omega_{press,outer} \right) \).  As expected from Keplerian motion, the inner spiral density wave angular frequency is significantly different from the outer spiral density wave angular frequency.  

\section{Numerical Simulation Codes and Results}
In this section, we describe two different numerical codes that generate accretion discs like those expected for non-magnetic CV DN.  In \S3.1, we summarize a grid code that generates accretion discs with an inner spiral density wave as shown in e.g., Bisikalo et al. (2003, 2004).  In this work, we present the $q=0.4$ numerical simulation using this code to see if lower mass ratio systems can also generate an inner spiral density wave.  In \S3.2, we summarize the SPH code that generates accretion discs with a denser inner ring if the disc is tilted at least a few degrees as shown in Montgomery (2009a).  

\subsection{3D, High Resolution, Gas Dynamical Grid Simulations} 
The details of the three-dimensional, high resolution gas dynamical grid code can be found in Bisikalo et al. (2003).  Here we provide a summary.  For this work, we consider a semi-detached binary consisting of a donor secondary star with mass $M_{2}=0.32M_{\odot}$ that fills its Roche lobe.  An accretor primary star with mass $M_{1}=0.8M_{\odot}$ is separated from the donor secondary by a distance $d=1.23R_{\odot}$. The modeling is carried out in a non-inertial reference frame rotating with the binary and on a regular three dimensional grid in Cartesian coordinates. The origin of the coordinate frame coincides with the center of mass of the donor secondary.

Since the problem is symmetrical about the equatorial plane, the modeling is carried out in half the space:  A symmetrical boundary condition is imposed at the corresponding boundary of the
computational domain.  Steady-state boundary conditions are specified at the remaining boundaries of the domain:  The density is $\rho_{b}=10^{-8}\rho_{L_1}$, where $\rho_{L_1} $ =1.5 x 10$^{-7} $ g $\cdot$ cm$^{-3}$ is the matter density at the inner Lagrange point $L_1$; the equilibrium temperature of the optically thin gas is $T=13,600 K$ (see Bisikalo et al. 2003); and the velocity is $\vec{v}_b=0$. The white dwarf accretor primary is specified to be a sphere with radius $R_{1}=0.015d$, and a free-inflow condition is imposed at its surface:  All the matter that falls in cells occupied by the white dwarf accretor is considered to have fallen onto the star.

The force field in the system is described by the Roche potential:
%%%%%%%%%%%%%%%%%EQUATIONS 32%%%%%%%%%%%%%%%%%%%
\begin{eqnarray}
\Phi & = & -\frac{GM_{2}}{\sqrt{x^2+y^2+z^2}}-\frac{GM_{1}}{\sqrt{(x-d)^2+y^2+z^2}}- \nonumber \\
        &     & \frac{\Omega^2}{2}\left[\left(x-d\frac{M_{1}}{M_{2}+M_{1}}\right)^2+y^2\right],
\label{eqroche}
\end{eqnarray}
%%%%%%%%%%%%%%%%%%%%%%%%%%%%%%%%%%%%
\noindent 
where $G$ is the gravitational constant and $\Omega$ is the rotational speed of the system. The gradient of the Roche potential is equal to zero at five libration points (the so-called Lagrange points):  $\nabla{\Phi}=0$. The equipotential surface that passes through the inner Lagrange point $L_1$ forms the Roche lobes of the system's components.

The size of the donor secondary, a red dwarf, is specified by the boundaries of its Roche lobe. The following boundary conditions have been selected at its surface: $\rho_{2}=\rho_{L_1}$, $T_{2}=3200$ K, $\emph{v}_{2}=\emph{v}_{L_1}\cdot\nabla{\Phi}/|\nabla{\Phi}|$. The speed at the inner Lagrange point $L_1$ is specified to be the local sonic speed: $|\emph{v}_{L_1}| = c_{s}$=6.6 kms$^{-1}$. The region within the donor secondary is excluded from the calculations.

We describe the flow structure in the binary using a system of gravitational gas dynamic equations, taking into account the radiative heating and cooling of the gas in the optically thin case:

\begin{equation}
\left\{
\begin{array}{l}
\frac{\partial\rho}{\partial t}+ \nabla \cdot (\rho \emph{v}) = 0\,, \\[3mm]
\frac{\partial\rho\emph{v}}{\partial t} + \nabla \cdot (\rho\emph{v}\otimes\emph{v}) + \nabla P = -\rho \nabla \Phi - 2 [{\Omega}\times\emph{v}]\rho\,,\\[3mm]
\frac{\partial\rho(\varepsilon+v^2/2)}{\partial t} + \nabla \cdot (\rho\emph{v})(\varepsilon + P/\rho + v^2/2) = \\
\indent \indent \indent -\rho\emph{v} \nabla \Phi + \rho^{2}m_p^{-2} \left[\Gamma(T,T_{a})-\Lambda(T)\right].
\end{array}
\right. \label{HDC}
\end{equation}

\noindent 
Here, $\rho$ is density, $\emph{v}=(u,v,w)$ is velocity, $P$ is pressure, $t$ is time, $\varepsilon$ the inner energy, $m_p$ is the mass of the proton, and $\Gamma(T,T_{a})$ and $\Lambda(T)$ are the radiative heating and cooling functions (see Bisikalo et al. 2003). The system of the gasdynamic equations is closed with the equation of state for an ideal gas, $P=(\gamma-1)\rho\varepsilon$, with an adiabatic index $\gamma = 5/3$ for a monatomic ideal gas.

We solve the system of equations using the Roe-Osher-Einfieldt method adapted for multiprocessor computers (Roe 1986; Chakravarthy \& Osher 1985; Einfeldt 1988). This technique makes it possible to obtain high-accuracy solutions of gas dynamical problems. The simulations are carried out until the system is in the steady-state accretion regime.

Figure 1 shows that a spiral arm can be generated in the steady state solution for the lower mass ratio $q=0.4$ numerical simulation where $M_{1}=0.8M_{\odot}$.  The cartesian coordinates (X,Y,Z) are in fractions of the separation distance $d$.  Although in this work we only show half the space, the full three-dimensional simulations also have the same spiral wave as shown in the purely hydrodynamic solutions by Bisikalo et al. (2008) and in the hydrodynamic solutions with magnetic fields by Zhilkin et al. (2010).  Therefore, the spiral density wave is not a consequence of symmetry and not due to magnetic fields.  

As shown in Figure 1, the spiral density wave does have a small projection relative the disc mid-plane.   Figure 2 shows snapshots of the disc over three-quarters of an orbit.  In their numerical simulations, Bisikalo et al. (e.g., 2003, 2004) find the wave to be a result of sheared elliptical particle orbits in the innermost annuli of the disc:  As the particles travel in elliptical orbits around the primary, where the primary is at one common focus to the elliptical orbits, the apastron points of the elliptical orbits are gravitationally torqued by the secondary to precess in the retrograde direction.  The closer an apastron point is to the secondary, the larger its semi-major axis, and the more torque that apastron point experiences.  Because velocity flow is very slow at these apastron points and because flow lines of the gas stream cannot cross, gas particles accumulate near these apastron points.  We find in this work that these accumulations have two sources:  gas particle migration from outer annuli and, as shown in Figure 10 of Bisikalo et al. (2005), the gas stream that overflows the disc edge and travels radially across the disc face to inner annuli.  The result of an over population of gas particles near apastron points is an elliptical spiral density wave in the innermost annuli.  Because apastron points are not uniformly spaced in angle due to differential rotation of the disc, accumulation of gas particles is not uniform near these points. The result is a variation in density along the wave.  The net of the two effects is an asymmetric, elliptical, tilted spiral density wave located in inner annuli as shown in Figure 1.  Because the center of mass is not in the center of the elliptical spiral and not in the orbital plane, the spiral density wave is subject to torques by the secondary and thus the spiral density wave retrogradely precesses.  We find for the q=0.4 numerical simulations in steady state that as the secondary orbits the center of mass, the secondary returns to its starting position (relative to the spiral density wave) at a period that is $\sim$3\% shorter than the orbital period.  This shorter period indicates that the spiral density wave is in retrograde precession.  In $\sim$ 33 orbits, the spiral density wave has made one full retrograde turn.  We note that the disc does not retrogradely precess.

From Figures 1 and 2, we find that the inner spiral density wave outer radius ranges from $\sim(0.1-0.2)$d.  However, the average outer radius is $\sim$0.16d, confirming our assumption that the inner spiral density wave is located near the inner ring seen in the SPH simulations  Montgomery (2009a).

%%FIGURE 1 %%%%%%%%%%%%%%%%%%%%%%%%
\begin{figure}
%\epsfxsize 3.3in
%\center{\epsfbox{Fig1.eps}}
{\caption{The inner spiral density wave is shown in cartesian coordinates as fractions of the separation distance $d$ for the q=0.4 numerical simulation.  Note the Z axis scaling is significantly smaller than the others, and only the upper half of the disc is shown.  The secondary is not shown.  However, the secondary is orbiting counterclockwise, and it is located at (X=0, Y=0, Z=0).  
\label{Figure 1.}}}
\end{figure}
%%%%%%%%%%%%%%%%%%%%%%%%%%%%%%%%%%
%%FIGURE 2 %%%%%%%%%%%%%%%%%%%%%%%%
\begin{figure}
%\epsfxsize 3.3in
%\center{\epsfbox{Fig2.eps}}
{\caption{The inner spiral density wave from Figure 1 is shown for orbits 30, 30.27, 30.49, and 30.74.  To convert into time, multiply one orbit by $P_{orb}$.  Cartesian coordinates are in fractions of the separation distance d.  The secondary is orbiting counterclockwise and is located at (X=0, Y=0, Z=0) in all panels.  
\label{Figure 2.}}}
\end{figure}
%%%%%%%%%%%%%%%%%%%%%%%%%%%%%%%%%%%

\subsection{3D, 100000 particle, SPH Simulations}
For the numerical simulations in Montgomery (2009b), we use Smoothed Particle Hydrodynamics (SPH) which is a Lagrangian method that models highly dynamical, astrophysical fluid flow as a set of interacting particles (see Monaghan 1992 for a review).  The code utilizes the simplest SPH form - constant and uniform smoothing length $h$ and constant and uniform mass particles.  Modifications to the original code include increasing particle number to 100,000 and adopting a Smith \& Dhillon (1998) secondary-to-primary mass-period relation.  The code is discussed further in Montgomery (2009b), however applicable information is also summarized here. 

The original code models hydrodynamics assuming an ideal gamma-law equation of state \( P= (\gamma-1) \rho u \) where $P$ is pressure, $\gamma$ is the adiabatic index, $\rho$ is density, and $u$ is specific internal energy.  The sound speed is  \( c_{s} = \sqrt{\gamma (\gamma -1) u}\).  The momentum and internal energy equations per unit mass are, respectively,

%%%%%%%%%%%%%%%%%%%EQUATION 25%%%%%%%%%%%%%%%%%%%%%%
\begin{equation}
\frac{d^{2}\mathbf{r}}{dt^{2}} = -\frac{\nabla P}{\rho} + \textbf{f}_{visc} - \frac{GM_{1}}{r_{1}^{3}} \textbf{r}_{1} - \frac{GM_{2}}{r_{2}^{3}} \textbf{r}_{2},
\end{equation}
%%%%%%%%%%%%%%%%%%%%%%%%%%%%%%%%%%%%%%%%%%%%%%%%

\noindent
and

%%%%%%%%%%%%%%%EQUATION 26%%%%%%%%%%%%%%%%%%%%%%%%
\begin{equation}
\frac{du}{dt} = -\frac{P}{\rho} \nabla \cdot \textbf{v} + \epsilon_{visc}
\end{equation}
%%%%%%%%%%%%%%%%%%%%%%%%%%%%%%%%%%%%%%%%%%%%%%

\noindent
in their most general form (Simpson 1995).  In these equations, ${\textbf{f}_{visc}}$ is the viscous force, $\epsilon_{visc}$ is the energy generation due to viscous dissipation, and $\textbf{r}_{1} $  and $\textbf{r}_{2} $ are the displacements from  stellar masses $M_{1}$ and $M_{2}$, respectively, that are given in units of solar masses.  The momentum and energy equations in SPH form for particles $i$ and $j$ are, respectively,

%%%%%%%%%%%%%%%%EQUATION 27%%%%%%%%%%%%%%%%%%%%%%%
\begin{eqnarray}
\frac{d^{2} \mathbf{r}_{i}}{dt^{2}} & = & -\sum_{j} m_{j}\left(\frac{P_{i}}{\rho_{i}^{2}}+\frac{P_{j}}{\rho_{j}^{2}}\right)(1+\Pi_{ij})\nabla_{i}W_{ij} - \nonumber \\
                          &   & \mbox{} \frac{GM_{1}}{r_{i1}^{3}} \textbf{r}_{i1} - \frac{GM_{2}}{r_{i2}^{3}} \textbf{r}_{i2}
\end{eqnarray}
%%%%%%%%%%%%%%%%%%%%%%%%%%%%%%%%%%%%%%%%%%%%%%

\noindent
and

%%%%%%%%%%%%%%%%EQUATION 28%%%%%%%%%%%%%%%%%%%%%%
\begin{equation}
\frac{du_{i}}{dt} = -\textbf{a}_{i} \cdot \textbf{v}_{i}
\end{equation}
%%%%%%%%%%%%%%%%%%%%%%%%%%%%%%%%%%%%%%%%%%%%%

\noindent
or, if the last equation fails in preventing negative internal energies,

%%%%%%%%%%%%%%%%%%%%EQUATION 29%%%%%%%%%%%%%%%%%%%
\begin{equation}
\frac{du_{i}}{dt} = \frac{P_{i}}{\rho_{i}^{2}} \sum_{j} m_{j} (1+\Pi_{ij})\textbf{v}_{ij} \cdot \nabla_{i} W_{ij}.
\end{equation}
%%%%%%%%%%%%%%%%%%%%%%%%%%%%%%%%%%%%%%%%%%%%%%%

\noindent
In these equations, $\Pi_{ij}$ is the Lattanzio et al. (1986) artificial viscosity 

%%%%%%%%%%%%%%%%%EQUATION 30%%%%%%%%%%%%%%%%%%%%%
\begin{equation}
\Pi_{ij}  =  \left\{ 
\begin{array}{ll}
 -\alpha \mu_{ij} + \beta \mu_{ij}^{2}	&  \qquad \mbox{$v_{ij} \cdot r_{ij}\leq 0$} \\
 0 & \qquad \mbox{otherwise}
\end{array}
\right.
\end{equation}
%%%%%%%%%%%%%%%%%%%%%%%%%%%%%%%%%%%%%%%%%%%%

\noindent
where

%%%%%%%%%%%%%%%%EQUATION 31%%%%%%%%%%%%%%%%%%%%%%
\begin{equation}
\mu_{ij} = \frac{h v_{ij} \cdot r_{ij}}{c_{s,ij}(r_{ij}^{2} + \eta^{2})}
\end{equation}
%%%%%%%%%%%%%%%%%%%%%%%%%%%%%%%%%%%%%%%%%%%%%

\noindent
and $c_{s,ij} = \frac{1}{2}(c_{s,i}+c_{s,j})$ is the average sound speed, $\textbf{v}_{ij}=v_{i}-v_{j}$, $r_{ij} = r_{i} - r_{j}$, and \( \eta^{2} = 0.01 h^{2} \) as shown in Simpson \& Wood (1998).  

For our numerical simulations, we choose \( \alpha = \beta = 0.5 \).  Our viscosity is approximately equivalent to a Shakura \& Sunyaev (1973) viscosity parametrisation (\(\nu = \alpha' c_{s} H \) where $H$ is the disc scale height) $\alpha$'=0.05.  As Smak (1999) estimates $\alpha' \sim $ 0.1 - 0.2 for DN systems in high viscosity states, our simulations are more for quiescent systems.  As only approaching particles feel the viscous force and since neither radiative transfer nor magnetic fields are included in this code, all energy dissipated by the artificial viscosity is transferred into changing the internal energies.  As radiative cooling is not included in this code, an adiabatic index $\gamma$=1.01 is incorporated to prevent the internal energies from becoming too large.  

By integrating the changes in the internal energies of all the particles over a specific time interval $n$, variations in the bolometric luminosity yield an approximate and artificially generated light curve 

%%%%%%%%%%%%%%%EQUATION 32%%%%%%%%%%%%%%%%%%%%%%%%%
\begin{equation}
L_{n} = \sum_{i} du_{i}^{n}.
\end{equation}
%%%%%%%%%%%%%%%%%%%%%%%%%%%%%%%%%%%%%%%%%%%%%%%
\noindent
All simulations start by injecting five particles per major time step, 200 major time steps per orbit, or a total of 1000 particles per orbit.  The injection velocity is based on an injection temperature $ T_{inj}$ =  4x10$^{4}$ K and is approximately the sound speed that has been scaled to system units.  The code adopts an approximate secondary mass-radius relation \( R_{2}=(M_{2}/M_{\odot})^{-13/15} R_{\odot} \) that applies for \( 0.08 \le \frac{M_{2}}{M_{\odot}} \le 1.0 \) (Warner 2003). In this relation, $R_{2}$ and $M_{2}$ are in solar radii and solar mass, respectively. 

A unique feature of this code is the conservation of disc particle number.  Any time a particle is 
accreted onto the secondary or primary mass or is lost from the system, a new particle is injected through the inner Lagrange point $L_{1}$.  We build a disc of 100,000 particles and maintain this number throughout the simulation.  After building the disc, we then allow the disc to evolve in the short term to a quasi-equilibrium state where only a few particles are injected per orbit.  At orbit 200, we artificially rotate the disc out of the orbital plane 5$^{\circ}$.  

For our simulations, we assume a primary mass $M_{1}$=0.8$M_{\odot}$ and we vary the secondary mass $M_{2}$ such that $0.35\le q \le 0.55$.  For example, if $q=0.4$ then $M_{2}$=0.32 $M_{\odot}$ and our simulation unit length scales to \( d \sim 1.23 R_{\odot} \) where we have used the secondary mass-radius relationship shown above and the Eggleton (1983) volume radius of the Roche lobe secondary

%%%%%%%%%%%%%%EQUATION 33%%%%%%%%%%%%%%%%%%%%%%%%%%%
\begin{equation}
\frac{R_{2}}{d} = \frac{0.49 q^{2/3}}{0.6 q^{2/3} + \ln(1 + q^{1/3})}.
\end{equation}
%%%%%%%%%%%%%%%%%%%%%%%%%%%%%%%%%%%%%%%%%%%%%%%%

\noindent
The Eggleton (1983) relation is good for all mass ratios, accurate to better than 1\%.  Taking the radius of the primary to be 6.9x10$^{8}$ cm, the scaled radius of the white dwarf is $R_{1}=0.0081d$.  

After building the disc, we allow the disc to evolve in the short term to a quasi-equilibrium state where particles are injected at the rate they are removed from the system by either being accreted onto the primary or the secondary or lost from the system.  The average net rate of accretion for a quasi-static disc is around 500 particles per orbit as shown in Montgomery (2009b).  As this net rate is approximately half that injected to build the disc, the steady state mass transfer rate reduces to $ \dot{m} \sim$ 1 x $10^{-10} M_{\odot}$/yr or a rate similar to an SU UMa in quiescence.  We estimate the density of the gas near $L_{1}$ to be $\rho \sim 10^{-10}$ g cm$^{-3}$ using \( m_{p}=(4/3)\pi h^{3} \rho \).   As we maintain 100,000 particles in the steady state disc, then the total mass of the disc is  $ M_{disc} $= 100,000 x $m_{p}$ = 3.5x10$^{22}$ g or \( M_{disc} \sim \) 2x10$^{-11} M_{\odot}$, a negligible value compared to the mass of either star.  If we assume the outer radius of the disc is \( R_{disc} \approx 2 r_{circ} \), or twice the circularization radius (Warner 2003)

%%%%%%%%%%%%%%%%%%%EQUATION 34%%%%%%%%%%%%%%%%%%%%%
\begin{equation}
\frac{r_{circ}}{a} = 0.0859 q^{-0.426}
\end{equation}
%%%%%%%%%%%%%%%%%%%%%%%%%%%%%%%%%%%%%%%%%%%%%%%

\noindent
where \( 0.05 \le q < 1 \), then for $q=0.4$ we find \( r_{circ} \sim 0.13 d \) and \( R_{disc} \sim 0.25 d \).   As \( \rho << M_{disc} R_{disc}^{-3} \), self gravity is neglected in these simulations.

As output to the code, we create an artificial light curve for each simulation (see Montgomery, 2009b). A non-linear least squares fit to each Fourier transform is computed and the standard deviation to twice the negative superhump frequency $2\nu_{-}$ is found.  This data in Table 1 is from Montgomery (2009b) and, along with the calculated negative superhump period,  the propagated error and the negative superhump period excess are also listed.  The units for period are hours whereas the units for frequency are orbits$^{-1}$ to be consistent with the Fourier transform figure.  To convert to real units, multiply one orbit by the orbital period.  

%%%%%%%%%%%%%%%%%TABLE 1%%%%%%%%%%
\begin{table*}
 \centering
 \begin{minipage}{200mm}
  \caption{Negative Superhump Simulation Data}
  \begin{tabular}{@{}lllllll@{}}
  		&\multicolumn{2}{c}{Frequency ($P_{orb}^{-1}$)}& & \multicolumn{2}{c}{Period (hr)}&\\
   		\cline{2-3} 
		\cline{5-6}
q & $2\nu_{-}$ & $\sigma_{2\nu_{-}}$ & $P_{orb}$ (hr) & $P_{-}$ & $\sigma_{P_{-}}$  & $\epsilon_{n}$ \\
 \hline
0.35   & 2.098 & 0.004 & 3.540 & 3.374 & 0.006 & 0.0467 \\
0.375  & 2.105 & 0.004 & 3.698 & 3.514 & 0.006 & 0.0498 \\
0.4    & 2.104 & 0.004 & 3.852 & 3.662 & 0.006 & 0.0494 \\
0.45   & 2.110 & 0.004 & 4.150 & 3.934 & 0.007 & 0.0521 \\
0.5    & 2.115 & 0.003 & 4.436 & 4.195 & 0.007 & 0.0544 \\
0.55   & 2.119 & 0.003 & 4.712 & 4.447 & 0.007 & 0.0562 \\
\hline
\end{tabular}
\end{minipage}
\end{table*}
%%%%%%%%%%%%%%%%%%%%%%%%%%%%%%%%%%%%%%%%%%%%%%%%%%%%%

\subsection{Grid Versus SPH Numerical Methods}
Both the SPH simulations and the grid simulations have gas particle migration from outer annuli to inner annuli of the disc and have gas stream overflow.  However, the SPH simulations have an imposed disc tilt whereas the grid simulations do not.  The SPH simulations generate an inner ring of higher density whereas the grid simulations generate an inner spiral density wave.  The SPH simulations have gas particles traveling in circular orbits in inner annuli whereas the grid simulations have  gas particles traveling in elliptical orbits in inner annuli.  Thus a wave is produced in the grid simulations but not the SPH simulations.

SPH is a viscous method that heavily relies upon particle interaction.  If particle neighbors are few then fine resolution is lost.  Grid simulations are also not immune to resolution:  To see detail like an inner spiral density wave, a fine grid is needed like the one used in this work.  SPH also heavily relies on artificial viscosity.  The smoothing effect of the artificial viscosity is known to hinder the production of spiral waves (see e.g., Lanzafame 2003).  Although both the total particle number and the artificial viscosity may contribute to particle orbit shape in inner annuli of SPH simulations, the latter is likely the leading cause to the lack of an inner spiral wave in the SPH simulations. 

\section[]{COMPARISONS WITH OBSERVATIONS}
Table 2 lists observational data including orbital periods; negative and positive superhump periods; and negative and positive superhump period excesses for several CV systems.  If errors are known, they are listed in parentheses.  We select long orbital period systems so that we may compare their observational results with our higher mass ratio numerical simulation and with our theoretical results.  In the table, NL is novalike, PS is permanent superhumper, and IP is intermediate polar.  

%%%%%%%%%%%%%%%%%%%%%%%%%Table 2%%%%%%%%%%%%%%%%%%%%%
\begin{table*}
 \centering
 \begin{minipage}{200mm}
   \tiny
  \caption{Positive and Negative Superhump Observational Data, Calculated Precessional Periods, and Period Excesses}
  \begin{tabular}{@{}llllllllll@{}}
   System & Type &q  & $P_{orb}(d)$ & $P_{p_{a}}(d)$ & $P_{p_{n}}(d)$  &  $P_{a}(d)$  &  $P_{n}(d)$  & $\epsilon_{a}$ & $\epsilon_{n}$ \\
 \hline
V592 Cas$^{a}$   & 1 &$0.19^{b,+0.1}_{-0.09}$ & 0.115063(1) & 1.950 & 4.111 & 0.12228(1) & 0.11193(5) & 0.0625(5) & 0.027\\
                                 &        &  $0.248^{c}$&              &             &            &           &        &            & \\
TT Ari$^{d, e}$ & 2 &$0.25^{f}$ & 0.13755040(17) & 1.762 & 3.930 & 0.1492(1) & 0.1329 & 0.0847(7) & 0.034\\
                                 &        &  $0.315^{c}$   &             &              &            &           &        &            &\\
 		              &        &  $0.19\pm0.04^{e}$   &             &              &            &           &        &            &\\
V603 Aql$^{g,h}$ & 3 &$0.24\pm0.05^{i}$& 0.1380(5) & 2.587 & 4.409 & 0.1460(7) & 0.1340 & 0.0572(51) & 0.030  \\
                                &        &  $0.23^{c}$   &             &              &            &           &        &            &\\
V503 Cyg$^{j}$ & 4 & $0.183^{c}$ & 0.07771(28) & 1.890 & 2.921 & 0.08104(7) & 0.75694 & 0.0430(27) & 0.026 \\
V1159 Ori$^{c,k,l}$ & 5 & $0.142^{c}$& 0.0621801(13) & 2.001 & 0.7236 & 0.06417(7) & 0.05743(14) & 0.0320(11) & 0.080 \\
AM CVn$^{m, n}$ & 3 & $0.18\pm0.01^{o}$ & 0.011906(1) & 0.5627 & 0.6948 & 0.012166(1) & 0.011706 & 0.0218(1) & 0.017 \\
TV Col$^{p, q}$ & 3,6 & $\sim0.33^{p}$ & 0.229167 & 1.805 & 3.973 & 0.2625 & 0.21667 & 0.15 & 0.0545 \\
CN Ori$^{r,s}$ & 6 & & 0.163199(7) &  & 0.1421 &  & 0.1595 &  & 0.022 \\
ER UMa$^{l, t}$ & 5 & $0.140^{c}$ & 0.06366(3) & 0.4180 & 1.255 & 0.0654(5) & 0.0589(7) & 0.0314(11) & 0.074 \\
V751 Cyg$^{u}$ & 2 & & 0.144464(1) &  & 3.806 &  & 0.1394(1) &  & 0.0353(2) \\
V442 Oph$^{v}$ & 8 & $0.79\pm0.27^{w}$ & 0.12433 &  & 4.420 &  & 0.12090(8) &  & 0.027 \\
V1974 Cyg$^{x, y}$ & 3 & $0.24\pm0.04^{x}$ & 0.08126(1) & 1.805 & 2.990 & 0.08509(8) & 0.07911 & 0.0471(10) & 0.027 \\
                               &        &  $0.197^{c}$   &             &              &            &           &        &            &\\
AH Men$^{y, z}$ & 8 & $0.326^{c}$ & 0.12721(6) & 1.561 & 4.306 & 0.1385(2) & 0.12356 & 0.0887(16) & 0.029 \\
DW UMa$^{c,v}$ & 8 &$0.33^{aa}$ &0.13661& 2.2573 & 5.038 & 0.14541 & 0.1330 & 0.0644 (20)& 0.026 \\
                               &        &  $0.255^{c}$   &             &              &            &           &        &            &\\
                               &        &  $>0.24^{bb}$   &             &              &            &           &        &            &\\
PX And$^{c,y,cc}$ & 8 & $0.329(11)^{c}$&0.146353(1)&1.776&4.228&0.1595(2)&0.1415&0.0898 &0.0331 \\
BH Lyn$^{c,dd}$& 8 & $0.45^{ee,+0.15}_{-0.10}$& 0.15575(1) & 2.128 & 3.413 & 0.16805 & 0.1490 & 0.079 & 0.0433 \\
                               &        &  $0.41\pm0.26^{y}$ &             &              &            &           &        &            &\\
                               &        &  $0.301(15)^{c}$     &             &              &            &           &        &            &\\
RX J1643$^{v}$ & 8 & & 0.120560(14) & & 3.917 & & 0.11696(8) & & 0.032(2) \\
RR Cha$^{ff}$ & 6 & & 0.1401 & 4.705&5.049 & 0.1444 & 0.1363 & 0.031 & 0.027 \\
AT Cnc$^{gg}$  & 3,7 &0.32 - 1.04 & 0.2011(6) & & 5.051    & &0.1934(8)     & & 0.0179(10)  \\ 
                               &       & &              &  & 11.03           &  &  0.1975(8)          & &  0.0383(10)     \\
IR Gem$^{hh, ii}$ & 4 & $0.154^{c}$ & 0.06840 & 1.390& 2.105& 0.07194& 0.0663& 0.052& 0.031\\
TX Col$^{jj}$ & 6 & $>0.33$ & 0.2375 & 1.204& 1.669&0.2958&0.2083&0.25&0.1229\\
SDSS J0407$^{kk}$& 4? & &0.17017(3)& &6.727& & 0.166(1)& & 0.024(1)\\
SDSS J2100$^{ll}$  & 4& &$\sim$0.0833&1.767&4.083&0.0875&0.0817 & 0.050 & 0.020\\
CAL 86$^{mm}$ & 4 &                               &0.06613 & & 1.403 & &0.06313    & &0.045\\
KR Aur$^{nn}$    &  2 &  $0.60\pm^{oo}$ & 0.1628  & & 4.489 & & 0.1571(2) & &0.0350(2)\\
HS 1813+6122$^{pp}$& 8 & & 0.1479 & & 3.166 & & 0.1413 &  & 0.0446 \\
V2574 Oph$^{qq}$ & 9 &  & 0.14773 & & 3.429 & & 0.14164 & & 0.412 \\
RX 1643+34$^{v}$ & 8  &  & 0.12050(14) &  & 3.917 & & 0.11696(8) & & 0.03000(8) \\
AY Psc$^{rr}$          &  1   &   0.45$^{ss,0.15}_{0.20}$  &0.21732(9)  &   &  & &   0.2057(1) &                       & 0.0535\\ 
                                    &                        &  &                          &   &           & & 0.2063(1)    &  &  0.0506     \\
                                    &                        &   &                         &   &           & & 0.2072(1)    &  & 0.0466      \\
BF Ara $^{tt}$         &  1  & $\sim$0.21&0.084176(21)& & & 0.08797(1)&0.082159(4)&0.0451(3) & 0.0244(2) \\
V1193 Ori$^{uu}$ &  3      & & 0.1430 &  &  &  & 0.1362 &  & 0.0476\\

\hline
\end{tabular}
\\$^{1}$NL, $^{2}$VY Scl/NL, $^{3}$PS/NL, $^{4}$SU UMa/DN, $^{5}$ER UMa/DN, $^{6}$IP/NL, $^{7}$Z Cam,  $^{8}$SW Sex/NL, $^{9}$Nova, $?$classification uncertain \\
$^{a}$Taylor et al. (1998), $^{b}$Huber et al. (1998), $^{c}$Patterson et al. (2005), $^{d}$Kraicheva et al. (1999), $^{e}$Wu et al. (2002), $^{f}$Warner (1995), \\
$^{g}$Borczyk et al. (2003), $^{h}$Patterson et al. (1997), $^{i}$Arenas (2000), $^{j}$Harvey et al. (1995),  $^{k}$Patterson et al. (1995), $^{l}$Thorstensen et al. \\
(1997),  $^{m}$Skillman et al. (1999), $^{n}$Nelemans et al. (2001), $^{o}$Roelofs et al. (2006) $^{p}$Retter et al. (2003), $^{q}$Hellier \& Buckley (1993), \\
$^{r}$Schoembs (1982), $^{s}$Barrera \& Vogt (1989), $^{t}$Gao et al. (1999), $^{u}$Patterson et al. (2001), $^{v}$RX J1643.7+3402 - Patterson et al. (2002), \\
$^{w}$Shafter \& Szkody (1983), $^{x}$Retter et al. (1997), $^{y}$Patterson (1998), $^{z}$Rodriguez-Gil et al. (2007b), $^{aa}$Biro (2000), $^{bb}$Araujo-Betancor \\
et al. (2003), $^{cc}$Boffin et al. (2003),  $^{dd}$Dhillon et al. (1992), $^{ee}$Hoard \& Szkody (1997), $^{ff}$ Woudt \& Warner (2002), $^{gg}$Kozhevnikov \\
(2004), $^{hh}$Li et al. (2004), $^{ii}$Fu et al. (2004), $^{jj}$Retter et al. (2005), $^{kk}$SDSS J040714.78-064425.1 - Ak et al. (2005), $^{ll}$SDSS J210014.12\\
+004446.0 - Tramposch et al. (2005), $^{mm}$Woudt et al. (2005), $^{nn}$Kozhevnikov (2007), $^{oo}$Kato et al. (2002), $^{pp}$Rodriguez-Gil et al. \\
(2007), $^{qq}$Kang et al. (2006), $^{rr}$Gulsecen et al. (2009), $^{ss}$Szkody \& Howell (1993), $^{tt}$Olech et al. (2007), $^{uu}$Ak et al. (2005b) \\
\end{minipage}
 \end{table*}
%%%%%%%%%%%%%%%%%%%%%%%%%%%%%%%%%%%%%%%%%%%%%%%%%

The disc geometry in this study is hypothetical as it is a combination of the disc geometries found in numerical simulations of Montgomery (2009b), Bisikalo et al. (2003, 2004), and this work.  The numerical simulations of Montgomery (2009b) show a tilted disc with a coplanar denser inner ring.  The numerical simulations of Bisikalo et al. (2003, 2004) and this work for $q=0.4$ show a coplanar disc with a dense inner spiral density wave that projects at a slight angle to the disc mid-plane.  Hence the hypothetical disc geometry for this study is a fully tilted disc around the line of nodes and the disc contains an inner spiral density wave that remains coplanar with the disc.  

Figure 3 shows a comparison of the numerical negative superhump period excess ($\triangle$) to the observational negative superhump period excess (*) values listed in Tables 1 and 2, respectively.  In this figure, the width of the symbol can be taken as the error in the orbital period and, to express mass ratio in terms of orbital period, we include the Smith \& Dhillon (1998) secondary mass-period relation.  As the geometry of the accretion discs for each observational system is unknown, we do not differentiate between different CV types with different symbols even though IPs probably have a significant absence of inner disc annuli, NLs could have discs with spiral density waves that cause the positive superhump as well as an inner spiral density wave or ring that causes the negative superhump, SU UMa's could have only the spiral density waves that cause the positive superhumps, etc.  Therefore, caution is advised when comparing the observational data with the numerical simulations as the scatter in the observational data is likely due to different geometries of a fluid disc (see Montgomery 2009a).  We also note that we have not taken into account other factors that may affect angular momentum and retrograde precession such as mass loss from winds, jets, mass loss to circumbinary rings, etc.

In Figure 3, we plot as a dash-dot line a differentially rotating, retrogradely precessing disc [i.e., from Equation (1)].  The dashed line represents a differentially rotating ring that can be found near the innermost annuli of the disc as discussed in Montgomery (2009a).  The dotted line represents Equation (31), the retrograde precession due to an inner spiral density wave that is located near the ring's radius.  As shown, the dotted line compares well with the dashed line.  Likewise, the net effect of a differentially rotating disc with a spiral density wave (the short dash line) compares well with the net effect of a differentially rotating disc with an attached, inner ring (the solid line).  In other words, the effects of pressure within the disc due to an innermost spiral density wave or the effects due to an attached innermost ring are necessary to better explain net retrograde precession in CV DN tilted, accretion discs should either be present in inner annuli.  

%%FIGURE 3 %%%%%%%%%%%%%%%%%%%%%%%%
\begin{figure}
%\epsfxsize 3.3in
%\center{\epsfbox{Fig3.eps}}
{\caption{Negative Superhump period excess as a function of orbital period.  Observational data are shown as stars and numerical data are shown as triangles.  The dashed line represents dynamical only contributions to retrograde precession.  The dotted line represents a spinning, tilted disc with an inner spiral density wave.  The short dash line combines the effects of a differentially rotating, tilted disc with the pressure effects from an spiral density wave.  The solid line combines the effects of a differentially rotating, tilted disc and an attached rotating innermost ring. Data are from Tables 1 and 2.  disc is tilted 5 degrees.
\label{Figure 3.}}}
\end{figure}
%%%%%%%%%%%%%%%%%%%%%%%%%%%%%%%%%%%

For the q=0.4 numerical simulation presented in this work, we find that the spiral density wave takes $\sim$33 orbits or $\sim$5.297d to retrogradely precess.  The negative superhump period is $\sim$3\% shorter than the orbital period or $\sim$3.736 hours.  The negative superhump period excess is $\sim$0.03.   

\section[]{DISCUSSION}
From the results, we can say that non-magnetic CV DN that show negative superhumps in their light curves can be described by a tilted, differentially rotating, precessing disc (i.e., the dynamical component) that has either a retrogradely precessing spiral density wave located in the innermost annuli or a retrogradely precessing ring that is superimposed on the innermost annuli of the disc.  Therefore, with the addition of either a retrogradely precessing ring or a retrogradely precessing spiral density wave, the geometry of the disc has changed as well as the net angular momentum and the net retrograde precession due to the change in the moment of inertia. Unknown is whether an inner spiral density wave or an inner denser ring is present.  However, both the ring and the spiral density wave have the same effect on the disc so long as both have nearly the same radial extent and are coplanar with the tilted disc.  If either is not coplanar with the disc, then differential precession would be present, effects not seen in non-magnetic CV DN systems.

We note in Figure 3 that we assume a Smith \& Dhillon (1998) secondary mass-period relation to obtain the negative superhump period excess as a function of orbital period.  By assuming this relation, the secondaries are taken to be main-sequence stars.  Those observations that agree with our numerical simulations and the solid and dashed lines may have tilted accretion discs with either an innermost precessing ring as discussed in Montgomery (2009a) or an innermost spiral density wave (this work).   

Bisikalo et al. (e.g., 2003, 2004) find that their numerically simulated inner spiral density wave retrogradely precesses yet the rest of the disc does not.  We confirm this result in the q=0.4 numerical simulation presented in this work.  Because this spiral density wave has its center of mass off-center and above the orbital plane of the disc, the spiral density wave is subject to torques by the secondary.  The net result is a retrogradely precessing spiral density wave.  The values for the negative superhump period, retrograde precession, and negative superhump period excess are long but not unreasonable when compared with the values in Table 2.  

\section{SUMMARY AND CONCLUSIONS}
Negative superhumps are found in observational systems with high mass ratios, long orbital periods, and high mass transfer rates.  In Montgomery (2009b), we find that the negative superhump can be due to extra light being emitted from inner disc annuli.  The extra light is modulated by gas stream overflow that strikes inner disc annuli and by gas particle migration into inner annuli as the secondary orbits.  For this result to be true, we find that the disc must be fully tilted around the line of nodes.  

In non-magnetic binary systems, discs that are tilted also retrogradely precess.  Patterson et al. (1993) suggest that retrograde precession in CV DN accretion discs may be by the same source that causes the Earth to retrogradely precess.  In Montgomery (2009a), we generate theoretical expressions for a disc that is misaligned with the orbital plane to test this suggestion, and we find that spinning, tilted, precessing accretion discs in non-magnetic binary systems can be described by a differentially rotating and precessing disc that has a precessing, denser ring superimposed on disc near the innermost annuli of the disc.  In this work, we consider a differentially rotating and precessing accretion disc that has an inner spiral density wave instead of a ring but having the same radial extent as the ring and in the same plane as the disc.  That is, in addition to dynamical effects, we consider pressure effects in our analytical expressions.  We adopt the theory used to describe pressure effects within CV DN discs that progradely precess.  

We compare our analytical expressions generated in this work with the analytical expressions generated in Montgomery (2009a) and the numerical simulations generated in Montgomery (2009b) for the mass ratio range \( 0.35 \le q \le 0.55 \).  In Montgomery (2009a), we find that non-magnetic, differentially rotating, tilted accretion discs can be described by a differentially rotating, retrogradely precessing accretion disc that has a superimposed retrogradely precessing ring in the innermost annuli of the disc.  We find in this work that these same non-magnetic, differentially rotating, tilted accretion discs can equally be described by a differentially rotating, retrogradely precessing accretion disc that has an innermost spiral density wave so long as the radial extent of the wave is like the radial extent of the ring.  Therefore, we conclude that pressure effects in retrogradely precessing systems would need to be included if negative superhumps are present in the light curve and if the fluid disc has an inner spiral density wave.

As part of this work, we generate a $q=0.4$ accretion disc with an inner spiral density wave using a 3D, high resolution, gas dynamical grid code.  The inner spiral density wave is not generated due to magnetic fields.  It is a consequence of elliptical particle orbits.  The inner spiral density wave projects at a small angle relative to the orbital plane.  The tilt is due to differing densities along the spiral density wave.   As a result of the center of mass being off-center and not in the orbital plane, the spiral density wave retrogradely precesses due to tidal torques, yet the disc does not.  The spiral density wave takes approximately 33 orbits to retrogradely precess, a value that is long but not unreasonable compared to the observations.  

\section*{Acknowledgments}
D. Bisikalo was supported by the Russian Foundation for Basic Research and by the Federal Agency of Sci. and Innovations.  We would also like to thank the anonymous referee.

}

\begin{thebibliography}{99}
\bibitem{b0} Ak T., Retter A., Liu A., \& Esenoglu H.H., 2005, PASA, 22, 105
\bibitem{b1} Ak T., Retter A., Liu A., \& Esenoglu H.H., 2005b, NewA, 11, 147
\bibitem{b2} Araugo-Betancor A., Knigge C., Long K.S., Hoard D.W., Szkody P., Rodgers B., Krisciunas K., Dhillon V.S., hynes R.I., Patterson J., \& Kemp J., 2003, ApJ, 583, 437 
\bibitem{b3} Arenas J., Catalan M.S., Augusteijn T., \& Retter A., 2000, MNRAS, 311, 135
\bibitem{b4} Barrett P., O'Donoghue D., \& Warner B., 1988, MNRAS, 233, 759
\bibitem{b5} Barrera L.H. \& Vogt N., 1989, A\&A, 220, 99
\bibitem{b6} Biro I.B., 2000, A\&A, 364, 573 
\bibitem{b7} Bisikalo D.V., Boyarchuk A.A., Kaigorodov P.V., \& Kuznetsov O.A., 2003, Astron. Rep. 47, 809
\bibitem{b8} Bisikalo D.V., Boyarchuk A.A., Kaygorodov P.V., Kuznetsov O.A., \& Matsuda T., 2004, ARep, 48, 449
\bibitem{b9} Bisikalo D.V., Kaigorodov P.V., Boyarchuk A.A., \& Kuznetsov O.A., 2005, ARep, 49, 701
\bibitem{b10} Bisikalo D.V., Kononov D.A., Kaigorodov P.V., Zhilkin A.G., Boyarchuk A.A., 2008, Astronomy Reports, Vol. 52, No. 4, p. 318
\bibitem{b11} Boffin H.M.J., Stanishev V., Kraicheva Z., \& Genkov V., 2003, ASPC, 292, 297
\bibitem{b12} Borczyk W., Schwarzenberg-Czerny A., Szkody P., 2003, A\&A, 405, 663
\bibitem{b13} Boyton P.E., Crosa L.M., Deeter J.E., 1980, ApJ, 237, 169
\bibitem{b14} Caproni A. \& Abraham Z., 2002, RMxAC, 14, 74
\bibitem{b15} Chakravarthy S.R. \& Osher S., 1985, AIAA, Paper 85-0363
\bibitem{b16} Dhillon V.S., Jones D.H.P., Marsh T. R., \& Smith R.C., 1992, MNRAS, 258, 225
\bibitem{b17} Eggleton P.P., 1983, ApJ, 268, 368
\bibitem{b18} Einfeldt B., 1988, SIAM J. Numer. Anal., 25, 294
\bibitem{b19} Foulkes S.B., Haswell C.A., \& Murray J.R., 2006, MNRAS, 366, 1399
\bibitem{b20} Frank J., King A.R., \& Raine D.J., 2002, Accretion Power in Astrophysics (3rd ed.; Cambridge:  Cambridge Univ. Press)
\bibitem{b21} Fu H., Li Z.-Y., Leung K.-C., Zhang Z.-S., Li Z.-L., \& Gaskell C.M., 2004, ChJAA, 4, 88 
\bibitem{b22} Gao W., Li Z., Wu X., Zhang Z., \& Li Y., 1999, ApJ, 527, 55
\bibitem{b23} Gulsecen H., Retter A., Liu A., Esenoglu H., 2009, NewA, 14, 330
\bibitem{b24} Harvey D., Skillman D.R., Patterson J., \& Ringwald F.A., 1995, PASP, 107, 551 
\bibitem{b25} Hellier \& Buckley D.A.H., 1993, MNRAS, 265, 766
\bibitem{b26} Hoard D.W. \& Szkody P., 1997, ApJ, 481, 433
\bibitem{b27} Huber M.E. Howell S.B., Ciardi D.R., \& Fried R., 1998, PASP, 110, 784
\bibitem{b28} Kang T.W., Retter A., Liu A., \& Richards M., 2006, AJ, 131, 1687
\bibitem{b29} Kato S., 1983, PASJ, 35, 249
\bibitem{b30} Kato S., 2001, PASJ, 53, 1
\bibitem{b31} Kato T., Ishioka R., \& Uemura M., 2002, PASJ, 54, 1033
\bibitem{b32} Katz J.I., 1973, Nature Phys. Sci., 246, 87
\bibitem{b33} Kim Y., Andronov I.L., Cha S.M., Chinarova L.L., \& Yoon J.N., 2009, A\&A, 496, 765
\bibitem{b34} Kozhevnikov V.P., 2004, A\&A, 419, 1035
\bibitem{b35} Kozhevnikov V.P., 2007, MNRAS, 378, 955
\bibitem{b36} Kraicheva Z., Stanishev V., Genkov V., \& Iliev L., 1999 A\&A, 351, 607
\bibitem{b37} Kumar S., 1986, MNRAS, 223, 225
\bibitem{b38} Lanzafame G., 2003, A\&A, 403, 593
\bibitem{b39} Larwood J., 1997, MNRAS, 290, 490
\bibitem{b40} Larwood J., 1998, MNRAS, 299, L32
\bibitem{b41} Larwood J., Nelson R.P., Papaloizou J.C.B., \& Terquem C., 1996, MNRAS, 282, 597
\bibitem{b42} Lasota J.-P., 2001, New Astronomy Review, 45, 449
\bibitem{b43} Lattanzio J.C., Monaghan J.J., Pongracic H., \& Schwarz M.P., 1986, J. Sci. Stat. Comput. 7, 591
\bibitem{b44} Li Z., Leung K.-C., \& Gaskell C.M., 2004, RMxAC, 21, 259
\bibitem{b45} Livio M., 1999, PhR, 311, 225
\bibitem{b46} Lubow S.H., 1991a, ApJ, 381, 259
\bibitem{b47} Lubow S.H., 1991b, ApJ, 381, 268
\bibitem{b48} Lubow S.H., 1992, ApJ, 398, 525
\bibitem{b49} Lubow S.H. \& Pringle J.E., 1993, ApJ, 409, 360
\bibitem{b50} Lubow S.H., Ogilvie G.I., \& Pringle J.E., 2002, MNRAS, 337, 706
\bibitem{b51} Meyer F. \& Meyer-Hofmeister E., 1981, A\&A, 104, L10
\bibitem{b52} Monaghan J.J., 1992, ARA\&A, 30, 543
\bibitem{b53} Montgomery M.M., 2001, MNRAS, 325, 761
\bibitem{b54} Montgomery M.M., 2004, Ph.D Thesis, Florida Institute of Technology
\bibitem{b55} Montgomery M.M., 2009a, ApJ, 705, 603
\bibitem{b56} Montgomery M.M., 2009b, MNRAS, 394, 1897
\bibitem{b57} Murray J.R. \& Armitage P.J., 1998, MNRAS, 300, 561
\bibitem{b58} Murray J.R., Truss M.R., \& Wynn G.A., 2002, ASPC, 261, 416 
\bibitem{b59} Nelemans G., Steeghs D., \& Groot P.J., 2001, MNRAS, 326, 621
\bibitem{b60} Olech A., Rutkowski A., \& Schwarzenberg-Czerny A., 2009arXiv0906.3964
\bibitem{b61} Olech A., Rutkowski A., \& Schwarzenberg-Czerny A., 2007, AcA, 57, 331
\bibitem{b62} Paczynski B., 1977, ApJ, 216, 822
\bibitem{b63} Papaloizou J. \& Pringle J.E., 1983, MNRAS, 202, 1181
\bibitem{b64} Papaloizou J. \& Lin D., 1995, ApJ, 438, 841
\bibitem{b65} Papaloizou J.C.B. \& Terquem C., 1995, MNRAS, 274, 987
\bibitem{b66} Patterson J., 1998, PASP, 110, 1132
\bibitem{b67} Patterson J., Thomas G., Skillman D.R., \& Diaz M., 1993, ApJS, 86, 235
\bibitem{b68} Patterson J., Jablonski F., Koen C., O'Donoghue D., \& Skillman D.R., 1995, PASP, 107, 1183
\bibitem{b69} Patterson J., Kemp J., Saad J., Skillman D., Harvey D., Fried R., Thorstensen .R., Ashley R., 1997, PASP, 109, 468
\bibitem{b70} Patterson J., Thorstensen J.R., Fried R., Skillman D.R., Cook L.M., \& Jensen L., 2001, PASP, 113, 72
\bibitem{b71} Patterson J., Fenton W.H., Thorstensen J.R., Harvey D.A., Skillman D.R., Fried R.E., Monard B., O'Donoghue D., Beshore E., Martin B., Niarchos P., VanMunster T., Foote J., Bolt G., Rea R., Cook L., Butterworth N., \& Wood M., 2002, PASP, 114, 1364
\bibitem{b72} Patterson J., Kemp J., Harvey D.A., Fried R.E., Rea R., Monard B., Cook L., Skillman D.R., Vanmunster T., Bolt G., Armstrong E., McCormick J., Krajci T., Jensen L., Gunn J., Butterworth N., Foote J., Bos M., Masi G., \& Warhurst P., 2005, PASP, 117, 1204
\bibitem{b73} Petterson J., 1977, ApJ, 216, 827
\bibitem{b74} Pringle J.E., 1996, MNRAS, 281, 357
\bibitem{b75} Pringle J.E., 1997, MNRAS, 292, 136 
\bibitem{b76} Pringle J.E., 1999, in Sellwood J.A., Goodman J., eds, ASP Conf. Ser. Vol. 160, Astrophysical Discs, Astron. Soc. Pac., San Francisco, p. 53
\bibitem{b77} Retter A., Leibowitz E.M., \& Ofek E.O., 1997, MNRAS, 286, 745
\bibitem{b78} Retter A., Hellier C., Augusteijn T., Naylor T., Bedding T.R., Bembrick C., McCormick J., \& Velthuis F., 2003, MNRAS, 340, 679
\bibitem{b79} Retter A., Liu A., \& Bos M., 2005, ASSL, 332, 251
\bibitem{b80} Rodriguez-Gil P. (and 17 others), 2007, MNRAS, 377, 1747
\bibitem{b81} Rodriguez-Gil P., Schmidtobreick L., \& Gansicke B.T., 2007b, MNRAS, 374, 1359
\bibitem{b82} Roe P.L., 1986, AnRFM, 18, 337
\bibitem{b83} Roelofs G.H.A., Groot P.J., Nelemans G., Marsh T.R., \& Steeghs D., 2006, MNRAS, 371, 1231
\bibitem{b84} Romero G.E., Chajet L., Abraham Z., \& Fan J.H., 2000, A\&A, 360, 57
\bibitem{b85} Schoembs R., 1982, A\&A, 115, 190
\bibitem{b86} Shafter A.W. \& Szkody P., 1983, PASP, 95, 509
\bibitem{b87} Shakura N.I. \& Sunyaev R.A., 1973, A\&A, 24, 337
\bibitem{b88} Simpson J.C., 1995, ApJ, 448, 822
\bibitem{b89} Simpson J.C. \& Wood M.A., 1998, ApJ, 506, 360
\bibitem{b90} Skillman D.R., Patterson J., Kemp J., Harvey D.A., Fried R.E., Retter A., Lipkin Y., Vanmunster T., 1999, PASP, 111, 1281
\bibitem{b91} Smak J., 1999, Acta Astronomica, 49, 391
\bibitem{b92} Smith A.J., Haswell C.A., Murray J.R., Truss M.R., \& Foulkes S.B., 2007, MNRAS, 378, 785
\bibitem{b93} Smith D.A. \& Dhillon V.S., 1998, MNRAS, 301, 767
\bibitem{b94} Szkody P. \& Howell S.B., 1993, ApJ, 403, 743
\bibitem{b95} Taylor C.J., Thorstensen J.R., Patterson J., Fried R.E., Vanmunster T., Harvey D.A., Skillman D.R., Jensen L., \& Shugarov S., 1998, PASP, 110, 1148
\bibitem{b96} Thorstensen J.R., Taylor C.J., Becker C.M., \& Remillard R.A., 1997, PASP, 109, 477
\bibitem{b97} Terquem C. \& Papaloizou J.C.B., 2000, A\&A, 360, 1031
\bibitem{b98} Tramposch J., Homer L., Szkody P., Henden A., Silvestri N.M., Yirak K., Fraser O.J., \& Brinkmann J., 2005, PASP, 117, 262
\bibitem{b99} Warner B., 1995, Cataclysmic Variable Stars (New York: Cambridge U. Press)
\bibitem{b100} Warner B., 2003, Cataclysmic Variable Stars (New York: Cambridge U. Press)
\bibitem{b101} Wijers R.A.M.J. \& Pringle J.E., 1999, MNRAS, 308, 207
\bibitem{b102} Wood M.A., Montgomery M.M., \& Simpson J.C., 2000, ApJL, 535, L39
\bibitem{b103} Woudt P.A. \& Warner B., 2002, MNRAS, 335, 44
\bibitem{b104} Woudt P.A., Warner B., \& Spark M., 2005, 364, 107
\bibitem{b105} Wu X., Li Z., Ding Y., Zhang Z., \& Li Z., 2002, ApJ, 569, 418
\bibitem{b106} Zhilkin A.G. \& Bisikalo D.V.
\end{thebibliography}
\end{document}